\documentclass{ws-procs9x6-cpt19}
\begin{document}

\newcommand{\refeq}[1]{(\ref{#1})}
\def\etal {{\it et al.}}

\title{Nonminimal Lorentz-Violating Extensions\\
of Gauge Field Theories
}

\author{Zonghao Li}

\address{Physics Department, Indiana University,\\
Bloomington, IN 47405, USA}

\begin{abstract}
A general method is presented
to build all gauge-invariant terms in gauge field theories,
including quantum electrodynamics and quantum chromodynamics.
It is applied to two experiments,
light-by-light scattering and deep inelastic scattering,
to extract first bounds on certain nonminimal coefficients 
for Lorentz violation.
\end{abstract}

\bodymatter

\section{Introduction}

Lorentz violation has been a popular topic in recent years
in the search for new physics beyond the Standard Model (SM).
The Standard-Model Extension (SME) 
developed by D.\ Colladay and V.A.\ Kosteleck\'y 
studies Lorentz violation in the context of effective field theory.\cite{ck97}
It includes all possible Lorentz-violating modifications
to the SM coupled to General Relativity
to describe all possible Lorentz-violating experimental signals.
All minimal terms (mass dimensions $d\leq4$)
have been established;\cite{ck97,ak04}
most nonminimal free-propagation terms 
have been established;\cite{km09}
and some low-dimension ($d\leq6$) interaction terms 
in quantum electrodynamics (QED) 
have been established.\cite{dk16}
However,
general Lorentz-violating terms in gauge field theories are still unknown.
Here,
we present a general method to build all Lorentz-violating terms
in gauge field theories
and apply these to two experiments,
light-by-light scattering and deep inelastic scattering (DIS),
to get first bounds on certain SME coefficients.
Related techniques can be applied in the gravity context.\cite{kl20}
The present contribution to the CPT'19 
proceedings is based on results in Ref.\ \refcite{kl19}.

\section{Theory}
\label{sec:theory}

The SME preserves gauge invariance,
so we need to find all gauge-invariant terms
to build general Lorentz-violating extensions
of gauge field theories.
The gauge-covariant operator
is a powerful tool in building gauge-invariant terms.
An operator $\mathcal{O}$ is called gauge covariant
if it transforms to $U\mathcal{O}U^\dagger$ 
under the gauge transformation,
where $U$ is a unitary representation of the gauge group $\mathcal{G}$,
and the fermion field $\psi$ transforms to $U\psi$ 
under the gauge transformation.
If $\mathcal{O}, \mathcal{O}_1, \mathcal{O}_2$ are gauge-covariant operators,
we can build gauge-invariant operators
by taking traces of them, 
$\textrm{Tr}(\mathcal{O})$,
or combining them with Dirac bispinors,
$\overline{(\mathcal{O}_1\psi)}(\mathcal{O}_2\psi)$.
Therefore,
we can first build gauge-covariant operators
and then get gauge-invariant operators.

A direct calculation shows that the gauge-covariant derivative $D_\mu$
and the gauge field strength tensor $F_{\alpha\beta}$
are gauge-covariant operators.
Moreover,
we find that any operator formed as a mixture of $D$ and $F$ is gauge covariant.
In principle,
we can construct gauge-invariant operators from all those operators.
However,
this would introduce a lot of redundancies 
because $F$ is related to the commutator of $D$ with itself.
Therefore,
we need to characterize those gauge-covariant operators
in terms of a set of standard bases with controlled or no redundancy.
The key result is that any operator formed as a mixture of $D$ and $F$
can be expressed as a linear combination of operators of the form
\begin{equation}
\label{eq:covariant}
(D_{(n_1)}F_{\beta_1\gamma_1})(D_{(n_2)}F_{\beta_2\gamma_2})\cdots
(D_{(n_m)}F_{\beta_m\gamma_m}) D_{(n_{m+1})},
\end{equation}
where $D_{(n)}=(1/n!)\sum D_{\alpha_1}D_{\alpha_2}\cdots D_{\alpha_n}$
is totally symmetrized with the summation performed 
over all permutations of $\alpha_1,\alpha_2,\cdots,\alpha_n$.
The basic idea behind Eq.\ \refeq{eq:covariant} 
is absorbing the symmetric parts in the totally symmetrized $D_{(n)}$
and the antisymmetric parts in $F$.
The detailed proof uses Young tableaux
and can be found in Ref.\ \refcite{kl19}.

We proceed to build general gauge-invariant terms from Eq.\ \refeq{eq:covariant}.
Both QED and quantum chromodynamics (QCD) 
are based on gauge field theories,
so the general Lorentz-violating extensions of QED and QCD can
be constructed.
We remark in passing 
that the extensions include both Lorentz-invariant 
and Lorentz-violating terms,
so we are actually building 
general gauge-invariant extensions of QED and QCD.
The reader is referred to Ref.\ \refcite{kl19}
for the details of the Lagrange densities.
In the next two sections,
we look at two experimental applications.

\section{Light-by-light scattering}

Light-by-light scattering is a nonlinear effect of the electromagnetic field,
which is hidden in the classical linear Maxwell equations
but can arise in QED via radiative loop corrections.
Experimental measurements of light-by-light scattering
can provide important tests of QED.
Since the cross section is tiny,
light-by-light scattering was directly measured only recently
at the LHC by the ATLAS collaboration.\cite{aetal17}
They measured ultraperipheral Pb+Pb collisions at $\sqrt{s_{NN}}=5.02\,$TeV.
By the equivalent-photon approximation,\cite{epa}
the collision of high-energy ultraperipheral heavy ions 
can be treated as collisions of photons from the heavy ions.

The QED extension built in the last section
can describe all possible deviations from the SM prediction
in light-by-light scattering experiments.
The dominant contribution comes from a $d=8$ term:
\begin{equation}
\label{eq:lbl}
\mathcal{L}_g^{(8)} \supset -\tfrac{1}{48}
k_F^{(8)\kappa\lambda\mu\nu\rho\sigma\tau\upsilon}
F_{\kappa\lambda}F_{\mu\nu}F_{\rho\sigma}F_{\tau\upsilon},
\end{equation}
where $k_F^{(8)\kappa\lambda\mu\nu\rho\sigma\tau\upsilon}$
are coefficients for Lorentz violation.
This term creates a new interaction vertex with four photon lines
and contributes to the light-by-light scattering at tree level.
Many possible Lorentz-violating signals can arise from this.
It produces new contributions to the total cross section
of light-by-light scattering in addition to the SM ones.
The SME coefficients are assumed to be approximately constant
in the Sun-centered frame,\cite{datatables}
so the experimental cross section can depend on the sidereal time
with the Earth rotating about its axis and revolving around the Sun.
The experimental results can also depend on the location and orientation of the laboratory.
The Lorentz-violating term can produce a new energy dependence 
for the differential cross section as well.

Due to statistical limitations of the data,
we compare here only the total cross sections 
to get bounds on the SME coefficients.
Future improvements in the experiment  
can lead to more detailed investigations of possible Lorentz-violating signals.
The LHC experiment measured the total cross section
as $70\pm24(\textrm{stat.})\pm17(\textrm{syst.})\,$nb.\cite{aetal17}
The theoretical SM prediction is $49\pm10\,$nb.\cite{ks10}
Comparing these two results gives bounds\cite{kl19} 
on 126 components of the coefficients $k_F^{(8)}$.
The bounds on the Lorentz-invariant and isotropic components of 
the coefficients $k_F^{(8)}$ are also extracted.
All these components are constrained to approximately $10^{-7}\,$GeV$^{-4}$.

\section{Deep inelastic scattering}

DIS provides key experimental support
for the existence of quarks and the predictions of QCD.
It is also an essential tool in the search for new physics beyond the SM
and can be used to test Lorentz symmetry.
The QCD+QED extension built via the method presented in Sec.\ \ref{sec:theory}
can describe all Lorentz-violating signals in DIS experiments.

The contributions from the minimal SME to DIS 
have been considered before.\cite{klv17}
Here, 
we focus on contributions from nonminimal terms.
Since most DIS experiments use unpolarized beams,
we consider spin-independent terms.
The leading-order spin-independent contribution from the nonminimal SME is 
\begin{equation}
\label{eq:dis}
\mathcal{L}_\psi^{(5)} \supset -\tfrac12
a_f^{(5)\mu\alpha\beta} \overline{\psi}_f \gamma_\mu
i D_{(\alpha} i D_{\beta)} \psi_f
+\textrm{h.c.},
\end{equation}
where the parentheses around the lower indices mean symmetrization 
on $\alpha$ and $\beta$ with a factor of $1/2$,
$f=u,d$ includes the dominant quark flavors,
and $a_f^{(5)\mu\alpha\beta}$ are $d=5$ coefficients for Lorentz violation.
This term modifies both the free propagation of fermions
and interactions among photons and fermions.
The cross section with the corrections 
can be found in Ref.\ \refcite{kl19}.

Based on the simulations in Ref.\ \refcite{klv17} for $c^{\mu\nu}$ coefficients,
we can estimate the bounds on $a^{(5)\mu\alpha\beta}$ 
to be around $10^{-7}-10^{-4}\,$GeV$^{-1}$.
We can also expect that the corrected DIS cross section depends
on up to the third-order harmonics of the sidereal-time variables
because the coefficients $a^{(5)\mu\alpha\beta}$ contain three indices.
The $a^{(5)\mu\alpha\beta}$ coefficients also provide CPT-odd contributions 
to DIS for protons and antiprotons.
Experimental measurements of those Lorentz-violating signals
can provide fruitful insight into new physics beyond the SM.

\section*{Acknowledgments}
This work was supported in part by the U.S.\ Department of Energy
and by the Indiana University Center for Spacetime Symmetries.

\end{document}